\documentclass[a4paper,fleqn,usenatbib,useAMS,twocolumn]{mnras}

\usepackage[T1]{fontenc}
\usepackage{ae,aecompl}

\usepackage{graphicx}	
\usepackage{amsmath}	
\usepackage{amssymb}	
\usepackage{longtable}
\usepackage{url}
\usepackage{hyperref}
\usepackage{supertabular}
\usepackage{ulem}

\usepackage{gensymb}
\usepackage{xspace}
\usepackage{amsmath}
\usepackage{amssymb}
\usepackage{lineno}
\usepackage{setspace}

\makeatletter

\newcommand{\Rmnum}[1]{\expandafter\@slowromancap\romannumeral #1@}
\makeatother

\newcommand{\secref}[1]{Section \ref{#1}}
\newcommand{\figref}[2]{Figure \ref{#1}{#2}\xspace}

\newcommand{\ARa}{AR 11309\xspace}
\newcommand{\ARb}{AR 11529\xspace}
\newcommand{\ARc}{AR 12738\xspace}

\title[Light Bridges]{Light Bridges Can Suppress the Formation of Coronal Loops}
\author[Y. H. Miao et al.]{
Yuhu Miao$^{1,2}$,
\newauthor{Libo Fu$^{1}$},
Xian Du$^{3}$,
Ding Yuan$^{1}$\thanks{E-mail:yuanding@hit.edu.cn},
Chaowei Jiang$^{1}$,
Jiangtao Su$^{2,4}$,
Mingyu Zhao$^{5}$,
\newauthor{Sergey Anfinogentov$^{6}$ }
\\
$^{1}$Institute of Space Science and Applied Technology, Harbin Institute of Technology, Shenzhen, Guangdong 518055, China\\
$^{2}$CAS Key Laboratory of Solar Activity, National Astronomical Observatories, Beijing 100012, China\\
$^{3}$College of Creative Design, Shenzhen Technology University, Guangdong 518118, China\\
$^{4}$School of Astronomy and Space Sciences, University of Chinese Academy of Sciences, 19 A Yuquan Road, Shijingshan District, Beijing \\ 100049, China\\
$^{5}$Yunnan Observatories, Chinese Academy of Sciences, Kunming 650011, China\\
$^{6}$Institute of Solar-Terrestrial Physics, 664033, Irkutsk, Russia}

\pubyear{2021}

\begin{document}
\label{firstpage}
\pagerange{\pageref{firstpage}--\pageref{lastpage}}
\maketitle

\begin{abstract}
A light bridge is a magnetic intrusion into a sunspot, it interacts with the main magnetic field and excites a variety of dynamical processes. In the letter, we studied magnetic connectivity between a light bridge and coronal loops rooted at the sunspot. We used the data of the Atmospheric Imaging Assembly onboard the {\sl Solar Dynamics Observatory} ({\sl SDO}) to study the features of sunspots with light bridges. It is found that if a light bridge anchors at the umbra-penumbra boundary, the coronal loops could not be formed around the anchoring point. If the a light bridge become detached from the penumbra, the coronal loop starts to form again. The vector magnetogram provided by the Helioseismic Magnetic Imager onboard {\sl SDO} shows that the anchoring region of a light bridge usually have an accompanying opposite minor-polarities. We conjugate that the magnetic field line could connect to these opposite polarities and form short-range magnetic loops, and therefore, coronal loops that extend to long-range could not be formed.  A model of light bridge is proposed to explain the magnetic connectivity between a light bridge and the coronal loops. This model could explain many physical processes associated with light bridges.
\end{abstract}




\section{Introduction}
\label{sec:intro}

A light bridge is an elongated bright structure that spans a sunspot or penetrate into it. They are commonly found in a nascent or decaying sunspot \citep{muller1979}. A light bridge is classified as a granular (or photospheric) light bridge, if convection is restored within it; or it is termed as a filamentary light bridge, if it has many thin filaments along its spine \citep{sobotka1997,katsukawa2007}. The granular bridges usually reveal photospheric dynamics and exhibit convection cells similar to the granules in the quiet Sun \citep{rouppe2010,sobotka2013}. It is suggested that convection probably plays a very important role in providing hotter gas to the light bridge \citep{lagg2014}. In the following text, we call the skeleton of a light bridge as the ``spine'', the longer edges in parallel with the spine are called ``flanks'', whereas two shorter anchors are named as ``ends''.

The magnetic field strength of a light bridge is commonly weaker than the surrounding
umbra of the sunspot \citep{sobotka2013}. However, \citet{castellanos2020} studied a counterexample, in which the light bridge of interest had a magnetic field strength exceeding 8 kG, this was the strongest magnetic field strength even measured in a sunspot light bridge. As a light bridge usually has magnetic field more inclined than the umbral field \citep{leka1997,katsukawa2007}, \citet{jurcak2006} proposed that a light bridge's magnetic field could have a canopy shape.  This configuration is driven by magneto-convection process as demonstrated with radiative MHD simulation by \citet{rempel2011} and \citet{siu2018}. Within the nearly horizontal magnetic field of a light bridge, the gas pressure of restored convection could become dominate and drags the magnetic field downward, and even results in reversed polarities at a light bridge at extreme cases \citep{lagg2014,felipe2016,zhangjw2018,guglie2017}.  Due to the strong discontinuity between the magnetic field of the light bridge and umbra, a strong electric current layer could be detected at the edges of a light bridge \citep{toriumi2015b}.

The magnetic configuration of a light bridge and umbra could excite many dynamic activities, such as jets \citep{asai2001,shen2011,shen2012,miao2018,miao2019b} and MHD waves \citep{yangsh2015,houyijun2016,zhangjw2017,felipe2017,lileping2018,miao2019a,yangxu2019,miao2020,miao2021,lifuyu2021}. \citet{louis2014} found that reconnection-excited jets usually originated from one flank of a light bridge and propagated towards one side. In order to explain the jet excitation and its asymmetric spatial distribution, \citet{yuan2016} proposed a three dimensional model for light bridge. The light bridge's magnetic field becomes nearly horizontal along the spine. Along the spine, the magnetic field has a helical component. Within this magnetic configuration, the magnetic field lines are aligned with the umbral field at one flank, whereas at the other flank the magnetic field lines of light bridge and umbra become anti-parallel. At this flank, magnetic reconnection could be triggered repetitively.  This model is supported by the high resolution ground observation of \citet{robustini2016} and \citet{tian2018}, wherein they detected $\lambda$-shaped jets straddling light bridges at its launching phase.

\citet{feng2020} find a potential correlation between the light bridge's anchoring position and the formation of coronal loops. The formation of coronal loops is a complex physical process associated with plasma heating and solar wind generation. \citet{wang2016,wang2019} found that at the footpoint of coronal loops a small and compact bipolar structure is usually formed. They also indicated that some of them have an inverted Y-shape. This configuration could trigger magnetic reconnections, which may provide forming material, moment and energy flux for large-scale magnetic structures. In the mean while, \citet{chitta2017} noted that coronal loops tend to root at regions with an accompanying mixed minority polarity.

In this letter, we aim to justify the magnetic connectivity between light bridge and coronal loops, and propose a novel model of light bridges in a sunspot. This letter is structured as follows: \secref{sec:obs} presents the observation and data analysis; and \secref{sec:results} present the case studies; the model of a light bridge and sunspot is proposed with an artistic drawing in \secref{sec:con}.

\section{Observations and data analysis}
\label{sec:obs}

To justify our initial proposition, we selected sample sunspots with light bridges and coronal loops. We had to focus on simple sunspot with a dominant polarity, so that the influence of strong opposite polarity was minimized, and the interaction between light bridge and coronal loops could be better traced. We used the 1700 \AA{} and 171 \AA{} channels of the Atmospheric Imaging Assembly \citep[AIA;][]{lemen2012}  on board the {\sl Solar Dynamic Observatory} ({\sl SDO}). The AIA 1700 \AA{} channel records plasma emission at ultraviolet (UV) continuum and is optimized for imaging observation of sunspots and light bridges; whereas the AIA 171 \AA{} channel captures the plasma emission at the extreme UV (EUV) bandpass, this channel is designed for EUV observation of coronal loops with temperature at about 0.8 MK. The LOS magnetogram was used to reveal the distribution of the magnetic polarities, the Space-weather HMI Active Region Patches \citep[SHARPs]{bobra2014} product was applied to display the vector magnetic field, and then transformed into standard heliographic spherical coordinates to match the AIA 171 \AA{} images. The data was provided by  the Helioseismic and Magnetic Imager \citep[HMI;][]{schou2012} on board {\sl SDO}.  The AIA has a pixel size of about $0\arcsec.6$, whereas the HMI's pixel corresponds to $0\arcsec.5$. The AIA and HMI data were calibrated with the standard procedure provided by the Solar Software (SSW)\footnote{\url{http://www.lmsal.com/solarsoft/ssw_install.html}}. The CCD readout noise and dark were removed from the images, and then the images were corrected with a flat-field and normalized with its exposure time.

\figref{fig:events}{} visualizes three sunspots with light bridges and associated coronal loop system, the sunspot samples were \ARa, \ARb and \ARc, observed on 6 October 2011, 29 July 2012, and 10 April 2019, respectively. The bottom row illustrates the the horizontal magnetic field vectors ($B_x,B_y$) overlaid on the background gray-scale image of vertical field component ($B_z$). In order to study the influence of light bridge over the coronal loops, we followed the evolution of sunspot \ARc (see \figref{fig:evol}), and investigated two comparative scenarios with or without a light bridge. In order to verify if this is common feature in sunspots, we check the {\sl SDO}/AIA observations from 2010 to 2020, and collected 66 such typical cases (see Table \ref{table}) that could validate our proposition.

\section{The correlation between light bridge and coronal loops}

\label{sec:results}

\figref{fig:events}{} shows three unipolar sunspots with light bridge, the associated active region loop systems, and magnetogram revealing the LOS and vector magnetic components.

\ARa was an alpha sunspot with negative polarity. Its umbra was segmented by a light bridge, which was oriented about $45\degree$ from the Solar-X axis in the Helioprojective-Cartesian coordinate, see \figref{fig:events}{(a2)}. \figref{fig:events}{(a1)} shows that coronal loops were not formed at the North-West region within the field-of-view, the magnetic field line could connect to local minor opposite polarities. The contour of LOS magnetic field shows that minor polarities opposite to that of the sunspot clustered close to the anchoring point of the light bridge, see \figref{fig:events}{(a3)}, these opposite polarities could be the other footpoints of small-scale magnetic field lines. \figref{fig:events}{(a4)} reveals that the horizontal magnetic field vector was aligned with the the spine of the light bridge. The magnetic vectors appear to point to a common source at the center of the bridge.

\ARb was also an almost circular alpha sunspot with positive polarity.  The light bridge in this sunspot was oriented about $-20\degree$ from the Solar-X coordinate as shown in \figref{fig:events}{(b2)}. The coronal loops were absent at the South-West periphery of \ARb, there was even little coronal EUV emission above the light bridge itself, see \figref{fig:events}{(b1)}. It appears that the formation of a light bridge made the coronal loops bifurcated into two bundles. Again, we also spotted the clusters of compact opposite polarities about a few arc seconds from an anchor of the light bridge, as shown in \figref{fig:events}{(b1) and (b3)}.

\ARc is an alpha sunspot with dominant negative polarity. The light bridge started from the umbra and was oriented towards the North-West direction. Large-scale coronal loop structures were rooted at sunspot \ARc, however, no coronal loop extended towards the North-West direction. A compact opposite polarity source was located close to the anchoring point of the light bridge.

\begin{figure}
\centering
\includegraphics[width=0.5\textwidth]{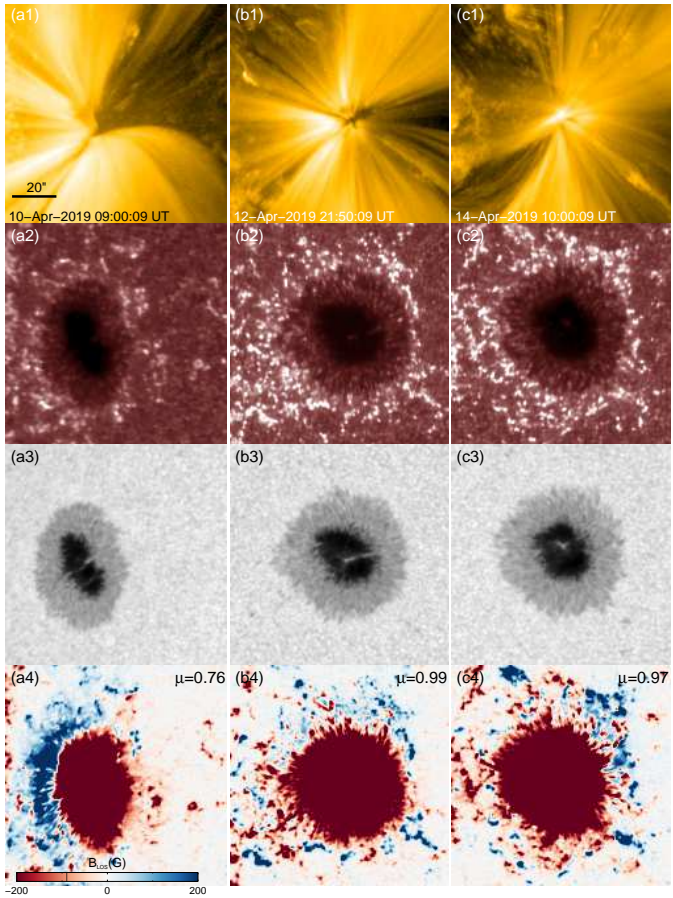}
\caption{From top to bottom rows: the AIA 171 \AA{} images (a), the AIA 1700 \AA{} images, the HMI continuum image, and the map of HMI LOS magnetic field. From left to right columns, the evolution of \ARc observed  on 10th, 12th and 14th April 2019, respectively.
\label{fig:evol}}
\end{figure}

\figref{fig:evol}{} draws the evolution of \ARc over five days from 10th to 14th April 2019. We could see clearly that on the 10 April, when the light bridge was anchored at the penumbra, no coronal loop was formed along the North-West direction (see \figref{fig:evol}{(a1)}). From 12 April and afterwards, the light bridge gradually became detached with the penumbra (see \figref{fig:evol}{(a3-c3)}), then coronal loop started to form again at the previous anchoring point (see \figref{fig:evol}{(a2-a3)}).

We present three sample sunspots with light bridges, it is clear within these samples that light bridge suppressed the formation of coronal loops at its anchor.   When the light bridge become detached with the penumbra,  coronal loops start to form again. This phenomenon is in fact very common in sunspots and active region loops. We inspect ten years observation of SDO/AIA and found 66 such cases, as listed in Table \ref{table}. We shall note that this list is not exclusive, and we have to focus on simple sunspots, as complex magnetic topology would not reveal this phenomenon.

\section{Conclusions and Discussions}
\label{sec:con}

In this letter, we aim to prove that light bridge could suppress the formation of coronal loops, and propose a novel model for a light bridge in order to explain the dynamics occurring at the small-scale magnetic structure within a sunspot, such as magneto-convection, magnetic reconnection, jet generation, and oscillations above light bridge, the formation and heating of coronal loops. Three typical examples are presented to display such phenomenon. The LOS and vector magnetic field maps reveal that close to the anchor of a light bridge, clusters of minor polarity opposite to that of the spot are usually found. It means the magnetic field line could start from the spot and connect to this local opposite polarities, rather than extend radially and form coronal loops. As short and compact magnetic loops have a different heating rate \citep{fisher1990,yokoyama1998}, the plasma could be heated to another temperature. We have checked the AIA channels that are sensitive to EUV emission of hotter plasmas (131\AA{}, 94\AA, and 335\AA{}), but no apparent compact hot loop systems were detected. The reason could be that this loop system could be heated chromospheric temperatures and could not be detected by AIA channels as studied by \citet{huangzhenghua2018,huangzhenghua2020}. Or, AIA hot channels are not sensitive enough to reveal this small structures.

A light bridge could be magnetically connected to the minor opposite polarities at the super-penumbra. The link could be developed by the magneto-convection process, as demonstrated in \citet{rempel2011} and \citet{toriumi2015a,toriumi2015b}. Therefore, the penumbra and light bridge could have a common submerged source, and this could explain why we could detect five-minute oscillation at these two structures and that they exhibit very identical features \citep{yuan2014,yuan2016,feng2020}. This type of magnetic connectivity is very similar to the moving magnetic features \citep[MMFs,][]{thomas2002}, which are indicators of submerged low-lying loops. However, the MMFs are magnetic flux pumped downwad  \citep{thomas2002}, whereas the minor opposite polarity in this study could be the rising magnetic structure driven by magneto-convection from \citep{rempel2011,siu2018}.

\begin{figure}
\centering
\includegraphics[width=0.5\textwidth]{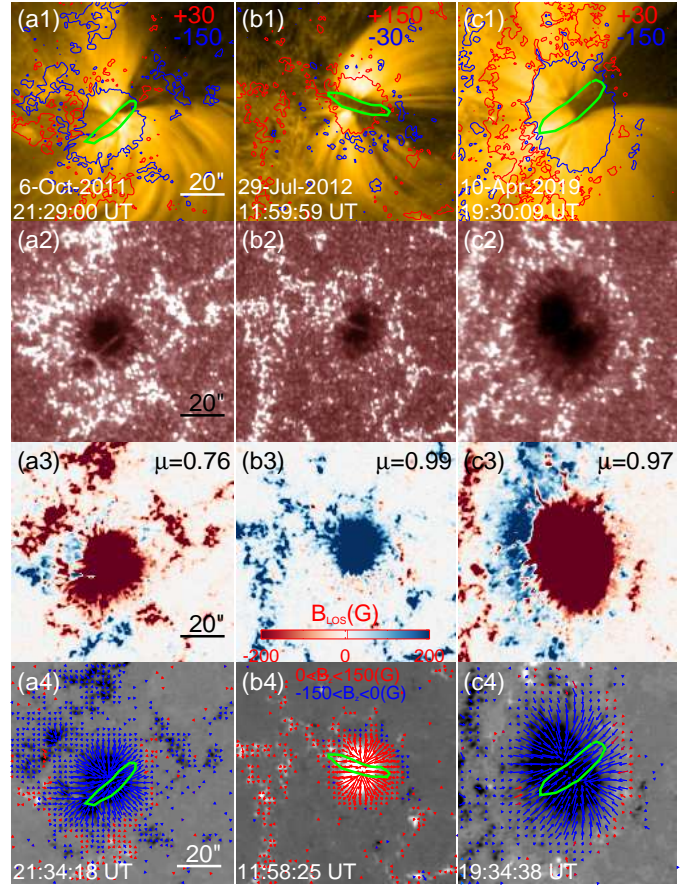}
\caption{Top row:  AIA 171 \AA{} images showing the corona loops structure, the red and blue contour reveals the LOS magnetic field component, the green contour shows the light bridge emission intensity in the AIA 1700 \AA{}. Second row: AIA 1700 \AA{} images illustrating the sunspots and the light bridges. Third row: HMI LOS magnetic field component. Bottom row: the horizontal field vector overlaid on the background
gray-scale image of vertical field component ($B_z$). The color of the arrows represents the polarity of vertical field component. From left to right columns, \ARa, \ARb and \ARc, observed on 6 October 2011, 29 July 2012, and 10 April 2019, respectively. We use $\mu=\cos{\theta}$ to label the effect of project, where the $\theta$ represents the longitude of the sunspot. \label{fig:events}}
\end{figure}

\begin{figure}
	\centering
	\includegraphics[width=0.5\textwidth]{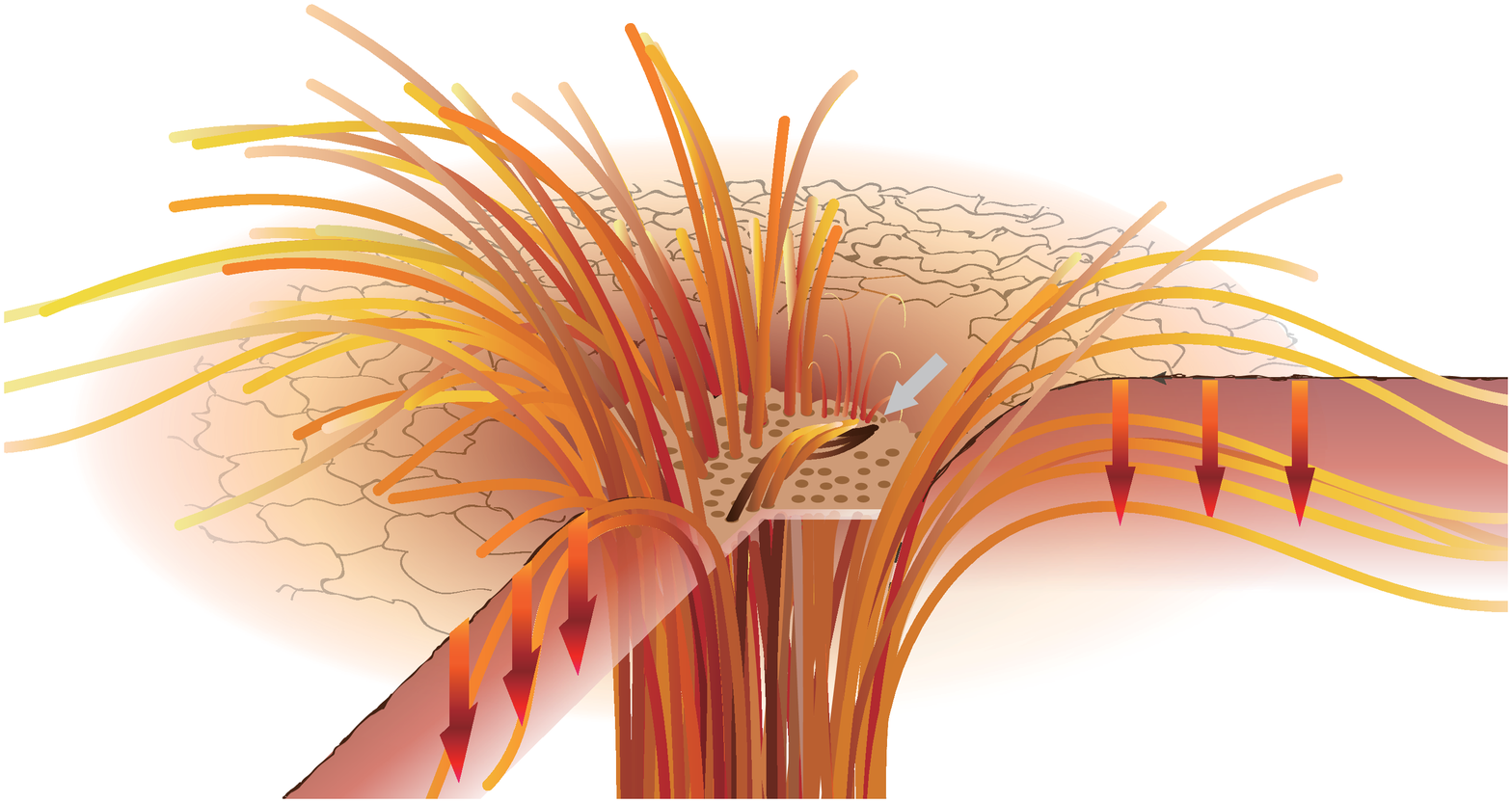}
	\caption{Artistic drawing showing the magnetic structure of a sunspot with a light bridge. The light bridge is more inclined to horizon and has a helical component along its spine. It connect to the opposite minor polarity at the umbra-penumbra boundary. Small-scale magnetic loop are also formed within this minor opposite polarity, no large extending loops are formed within that region. The anchoring region is indicated by a gray arrow.\label{fig:cartoon}}
\end{figure}

In order to explain connectivity of the active region loops and light bridge, a three dimensional model is presented with an artistic drawing as shown in Figure \ref{fig:cartoon}. The magnetic field is inclined more horizontally and bear twist along its spine. The twisted magnetic field of a light bridge was initially proposed by \citet{yuan2016}, it explains the excitation of $\lambda$-shaped jets \citep{tian2018,robustini2016} and the spatial asymmetry of jet origin and direction of jet propagation \citep{louis2014}. The magnetic field line of a light bridge anchors at another polarity at the supra-penumbra, this opposite polarity drags down the magnetic field lines that used to be corona loops. This model explains why coronal loops are not formed close to the anchor of a light bridge. This is very important in the comprehension of corona plasma confinement and the generation of solar wind, and other fundamental question of coronal loop heating. We should also note that a light bridge could have more than one anchor. The the horizontal magnetic field vector in \figref{fig:events}{} reveals that a light bridge could have two branches, meaning two anchors share one polarity at the umbra. If a light bridge have two anchors, it means that the magnetic flux bifurcate and connect to the opposite polarities at two anchors. This could explain in some cases at both ends of a light bridge, we could detect minor opposite polarities. The list is limited to simple sunspots, because the phenomenon does not show up in spots with complex magnetic topology. We also have to note that we do not prove that this phenomenon always takes place in simple sunspots.

\begin{table}
	\centering
	\caption{Selected sunspots with light bridge that suppress the formation of active region loop}
	\label{table}
	\begin{tabular}{cccccccc} 
		\hline
		{\bf Universal Time} & {\bf Active region} & {\bf Coordinates}\\
		\hline
    2010-09-21 20:00 & 11108 & $-103\arcsec, -570\arcsec$ \\

    2011-05-31 02:00 & 11226 & $-535\arcsec, -320\arcsec$  \\

    2011-07-17 16:00 & 11251 & $+55\arcsec, +195\arcsec$  \\

    2011-10-06 21:30 & 11309 & $-250\arcsec, +285\arcsec$  \\

    2012-06-12 17:30 & 11504 & $-415\arcsec, -300\arcsec$  \\

    2012-06-12 23:33 & 11507 & $-135\arcsec, -455\arcsec$  \\

    2012-07-12 12:06 & 11521 & $+280\arcsec, -413\arcsec$    \\

    2012-07-29 12:00 & 11529 & $+36\arcsec, -270\arcsec$   \\

    2012-07-31 23:10 & 11532 & $-10\arcsec, -410\arcsec$   \\

    2012-08-13 07:00 & 11543 & $-20\arcsec, +250\arcsec$   \\

    2012-08-25 12:00 & 11554 & $-60\arcsec, +140\arcsec$   \\

    2012-09-22 02:00 & 11575 & $-520\arcsec, +25\arcsec$   \\

    2012-10-24 03:00 & 11596 & $-25\arcsec, +50\arcsec$   \\

    2013-01-29 22:30 & 11663 & $-45\arcsec, -70\arcsec$   \\

    2013-03-29 12:00 & 11704 & $+12\arcsec, +360\arcsec$   \\

    2013-09-02 16:00 & 11836 & $+95\arcsec, +70\arcsec$  \\

    2013-09-20 00:00 & 11843 & $+370\arcsec, -90\arcsec$  \\

    2013-09-20 18:59 & 11849 & $+353\arcsec, +250\arcsec$  \\

    2013-09-21 16:59 & 11846 & $-410\arcsec, -400\arcsec$   \\

    2013-10-08 16:00 & 11857 & $+100\arcsec, -225\arcsec$  \\

    2013-10-13 23:33 & 11861 & $+240\arcsec, -230\arcsec$  \\

    2013-10-19 22:00 & 11872 & $-260\arcsec, -360\arcsec$   \\

    2013-10-20 22:00 & 11877 & $-735\arcsec, -255\arcsec$   \\

    2013-12-11 15:30 & 11917 & $-325\arcsec, -258\arcsec$   \\

    2013-12-20 15:30 & 11928 & $+515\arcsec, -255\arcsec$   \\

    2013-12-20 23:10 & 11930 & $-115\arcsec, -125\arcsec$  \\

    2014-01-20 01:00 & 11960 & $-815\arcsec, -230\arcsec$    \\

    2014-01-23 00:00 & 11960 & $-360\arcsec, -185\arcsec$   \\

    2014-02-04 01:16 & 11968 & $+60\arcsec, +260\arcsec$     \\

    2014-03-16 20:00 & 12005 & $-350\arcsec, +310\arcsec$  \\

    2014-05-01 05:08 & 12049 & $-550\arcsec, -35\arcsec$    \\

    2014-05-03 05:15 & 12047 & $+515\arcsec, -215\arcsec$    \\

    2014-05-18 20:00 & 12063 & $+150\arcsec, +200\arcsec$    \\

    2014-07-02 22:30 & 12107 & $-570\arcsec, -340\arcsec$        \\

    2014-07-24 03:00 & 12121 & $-760\arcsec, +75\arcsec$         \\

    2014-08-13 15:00 & 12139 & $-650\arcsec, +150\arcsec$        \\

    2014-08-17 05:00 & 12141 & $-440\arcsec, +165\arcsec$        \\

    2014-08-22 12:00 & 12146 & $-50\arcsec, +40\arcsec$        \\

    2014-08-25 12:00 & 12149 & $-380\arcsec, +60\arcsec$         \\

    2014-08-30 12:00 & 12151 & $+240\arcsec, -230\arcsec$       \\

    2014-09-20 11:30 & 12171 & $-795\arcsec, -205\arcsec$       \\

    2014-10-01 08:00 & 12178 & $-435\arcsec, -125\arcsec$       \\

    2014-10-23 01:00 & 12192 & $-195\arcsec, -330\arcsec$      \\

    2014-10-23 01:00 & 12193 & $+540\arcsec, +10\arcsec$       \\

    2014-10-29 08:00 & 12195 & $+215\arcsec, +50\arcsec$       \\

    2014-11-22 15:00 & 12216 & $-730\arcsec, -260\arcsec$       \\

    2014-12-10 11:00 & 12227 & $+410\arcsec, -50\arcsec$        \\

    2014-12-27 03:00 & 12246 & $-140\arcsec, +325\arcsec$       \\

    2015-03-27 22:00 & 12305 & $+175\arcsec, -55\arcsec$   \\

    2015-05-09 03:00 & 12339 & $-410\arcsec, +250\arcsec$   \\

    2015-08-27 17:00 & 12403 & $+660\arcsec, -355\arcsec$  \\

    2015-12-20 05:15 & 12470 & $+250\arcsec, +228\arcsec$   \\

    2016-01-11 20:07 & 12480 & $-58\arcsec, +118\arcsec$   \\

    2016-02-08 07:29 & 12494 & $+480\arcsec, -105\arcsec$  \\

    2016-04-13 22:58 & 12529 & $+20\arcsec, +275\arcsec$  \\

    2016-05-07 22:34 & 12542 & $-481\arcsec, +246\arcsec$  \\

    2016-06-12 07:30 & 12553 & $-775\arcsec, -120\arcsec$  \\

    2016-07-12 15:35 & 12564 & $-416\arcsec, +100\arcsec$  \\

    2016-09-07 15:30 & 12585 & $+368\arcsec, +30\arcsec$  \\

    2017-01-14 22:00 & 12625 & $-698\arcsec, +85\arcsec$  \\

    2017-01-14 23:30 & 12626 & $-796\arcsec, +178\arcsec$   \\

    2018-05-29 15:00 & 12712 & $-200\arcsec, +290\arcsec$    \\

    2019-04-10 19:20 & 12738 & $-556\arcsec, +176\arcsec$      \\

    2019-05-09 02:53 & 12740 & $-211\arcsec, +188\arcsec$     \\

    2019-05-11 21:19 & 12741 & $-210\arcsec, +137\arcsec$     \\

    2020-06-08 02:00 & 12765 & $-266\arcsec, -400\arcsec$     \\
		\hline\\

	\end{tabular}

\end{table}

\section{Acknowledgements}

Y.H.M. and D.Y. is supported by the National Natural Science Foundation of China (NSFC, 11803005,12111530078, 41731067), the Shenzhen Technology Project (JCYJ20180306172239618, GXWD20201230155427003-20200804151658001), the Shenzhen Science and Technology Program (Group No.KQTD20180410161218820), the China Postdoctoral Science Foundation (2020M681085) and the Open Research Program of the Key Laboratory of Solar Activity of Chinese Academy of Sciences (KLSA202110). C.W.J. is supported by the NSFC 41822404 and Shenzhen Technology Project JCYJ20190806142609035. M.Y.Z. is supported by the NSFC 11973086.

\section{Data availability}

The data underlying this article were accessed from the {\sl Solar Dynamics Observatory} ({\sl SDO})(\url{http://jsoc.stanford.edu/}). The derived data generated in this research will be shared on reasonable request to the corresponding author.


\end{document}